# Controlling Cherenkov threshold with nonlocality

Hao Hu[1,†], Xiao Lin[2,*,†], Jingjing Zhang[3], Dongjue Liu[1], Patrice Genevet[4], Baile Zhang[2,5,*] and Yu Luo[1,*]

[1]School of Electrical and Electronic Engineering, Nanyang Technological University, Nanyang Avenue, Singapore 639798, Singapore
[2]Division of Physics and Applied Physics, School of Physical and Mathematical Sciences, Nanyang Technological University, Singapore 637371, Singaporew
[3]State Key Laboratory of Millimeter Waves, Southeast University, Nanjing 210096, China
[4]CNRS, CRHEA, Universite Cote d'Azur, rue Bernard Gregory, 06560, Sophia Antipolis, Valbonne, France
[5]Centre for Disruptive Photonic Technologies, Nanyang Technological University, Singapore 637371, Singapore
*Corresponding author. E-mail: xiaolinbnwj@ntu.edu.sg; blzhang@ntu.edu.sg; luoyu@ntu.edu.sg
[†]These authors contributed equally to this work

**Cherenkov radiation is generally believed to be threshold-free in hyperbolic metamaterials owing to the extremely large photonic density of states in classical local framework. While recent advances in nonlocal and quantum effects extend our understanding of light-matter interactions in metallic nanostructures, the influence of nonlocality on threshold-free Cherenkov radiation still remains elusive. Here we theoretically demonstrate that the nonlocality provides an indispensable way to flexibly engineer Cherenkov thresholds in metallodielectric layered structures. Particularly, the nonlocality results in a lower-bound velocity cutoff, whose value is comparable to the electron Fermi velocity. Surprisingly, this lower-bound threshold can be significantly smaller than the classically predicted one if the metamaterial works around epsilon-near-zero frequencies. The capability to control Cherenkov thresholds opens numerous prospects for practical applications of Cherenkov radiation, in particular, for integrated free-electron radiation sources.**

**INTRODUCTION**

Cherenkov radiation was firstly discovered in experiments by P. A. Cherenkov in 1934 (*1*), under the supervision of S. I. Vavilov, and later theoretically interpreted by I. Frank and I. Tamm in 1937 (*2*). According to Frank and Tamm's theory, Cherenkov radiation arises when the velocity of charged particles exceeds a finite velocity threshold known as the Cherenkov threshold $v_{\text{th}}$. It is determined by the phase velocity of light in the surrounding medium, i.e., $v_{\text{th}} = c/n$, where $c$ is the light speed in free space and $n$ is the refractive index of the background medium. It is generally believed in the early twentieth century that photon emission from a uniformly moving charge is prohibited. The discovery of Cherenkov radiation changed such a belief. Thus, understanding the threshold behaviors of Cherenkov radiation is of paramount importance to the development of fundamental physics (*3-8*).

Cherenkov threshold is also an important parameter to many practical applications including elementary particle detectors, dosimetry, medical imaging and therapy, etc (*9-13*). As a typical example, the threshold determines the sensitivity of Cherenkov detectors for the identification of high-energy particles over a wide momentum range. To be specific, a higher sensitivity of the detector requires a higher Cherenkov threshold (i.e., $v_{\text{th}} \rightarrow c$), and hence a smaller refractive index of the background material. As a result, the design of Cherenkov detectors generally demands materials with the refractive index close to unity (e.g., gases, aerogels) in order to identify charged particles in the multi-giga-electron-volt (GeV) range (*14-16*).

The capability to flexibly engineer the Cherenkov threshold could benefit many applications mentioned above. For instance, decreasing the Cherenkov threshold enables the on-chip application of low-energy free electrons in compact and integrated light sources (*17-21*). However, owing to the limited value of the refractive index attainable with natural occurring materials (e.g., $n < 4$ for transparent dielectrics at the visible regime), the kinetic energy of swift electrons used in these light sources is generally above 1 mega-eV (MeV). To address this problem, artificial metallic nanostructures, including plasmonic waveguides and hyperbolic metamaterials, have been proposed (*22-28*). With a proper design, the effective refractive index of eigenmodes in these structures could go to infinity (i.e., $n \to \infty$) under the classical local approximation. Consequently, Cherenkov radiation can be made threshold-free under the local approximation, i.e. $v_{\text{th}} = c/n \to 0$. Following such theoretical proposals, a recent experiment reported the lowest Cherenkov threshold to date, i.e., $v_{\text{th}} = 0.03c$ (*29*), in hyperbolic metamaterials.

Further reducing the Cherenkov threshold beyond the results in Ref. 29 has remained difficult, probably because of the elusive role of nonlocality which is generally ignored in previous theoretical modellings. Although the nonlocality is known to modify dramatically the electromagnetic responses of metallic nanostructures (*30-35*), its influence on the threshold behaviors of Cherenkov radiation still remains unexplored.

Here we address this issue by systematically investigating the nonlocal effects on Cherenkov radiation in metallodielectric layered structures. These structures have applied widely to construct hyperbolic metamaterials. We find that the interplay between the spatial dispersion from the structural periodicity and the nonlocal electron screening in metals can strongly modify the threshold behaviors of Cherenkov radiation. First, the nonlocality always enables the Cherenkov threshold to be larger than the hydrodynamical velocity $\beta$ of plasma pressure waves in metals, namely $v_{\text{th}} \geq \beta$, for metallic nanostructures including hyperbolic metamaterials. Such a finding is in stark contrast to the previous prediction of the zero-value Cherenkov threshold in hyperbolic metamaterials under the local approximation. Second, around the epsilon-near-zero frequency, the nonlocality may largely *decrease* the value of the Cherenkov threshold. Such a counterintuitive finding originates from the appearance of longitudinal modes inside metallodielectric layered structures. Note that all the features of longitudinal modes cannot be captured by the local approximation. Moreover, our findings also indicate that the nonlocality may provide novel degree of freedom in controlling the interaction between free electrons and photonic systems, and it can be exploited for the flexible engineering of Cherenkov thresholds.

## RESULTS

**Theoretical description of Cherenkov radiation in local and nonlocal framework**

The basic structural setup is shown in Fig. 1A. A swift electron moves with a velocity of $\bar{v} = \hat{z}v$ in free space, along a trajectory close to the top surface of the metallodielectric layered structure. The unit cell of the layered structure has a pitch of $P = d_1 + d_2$, where $d_1$ and $d_2$ are the thicknesses of the metal (e.g., silver) and dielectric (silicon nitride) slabs, respectively. Here we set $d_1/P = 0.4$ and $P \ll \lambda$, where $\lambda \in [0.1\ 10]$ μm is the interested range of wavelength. Under the local approximation, the metallodielectric layered structure is treated as a homogeneous uniaxial material with a relative permittivity of $\bar{\bar{\varepsilon}}_r = [\varepsilon_{||}, \varepsilon_\perp, \varepsilon_{||}]$, where $\varepsilon_{||}$ is the in-plane permittivity and $\varepsilon_\perp$ is the out-of-plane permittivity. In particular, the structure in Fig. 1A is the hyperbolic metamaterial with $\text{Re}(\varepsilon_{||}) \cdot \text{Re}(\varepsilon_\perp) < 0$ if $\lambda > 0.4$ μm [Fig. S1].

On the other hand, when considering the nonlocality from the structural periodicity, the layered structure is used in the calculation by employing the transfer matrix method (*36*). When further considering the nonlocality from metals in the realistic layered structure, the hydrodynamic model is adopted to describe the nonlocal electron screening in metals. The nonlocal electron screening refers to the spatial spreading of surface charge densities over boundaries of metals, originating from the quantum repulsion of electrons or the electron gas pressure (*33, 37, 38*). The oscillation of free electron gas due to the pressure gives rise to a longitudinal response of metals. The corresponding longitudinal permittivity is wavevector-dependent, that is $\varepsilon_L(\omega, k_L) = 1 - \frac{\omega_p^2}{\omega^2 + i\gamma\omega - \beta^2 k_L^2}$ (*39*), where $\omega_p$ is the plasma frequency, $\gamma$ is the damping rate, and $k_L$ is the wavevector of longitudinal electric fields. The phenomenological nonlocal parameter $\beta$ is proportional to $\sqrt{E_F/m_e}$ or $v_F$ (*40*), where $E_F$ is the Fermi energy, $v_F$ is the Fermi velocity and $m_e$ is the effective electron mass. To consider the nonlocal response of metals, an additional boundary condition is needed to solve the Maxwell's equations; see details in Supplementary Section 6.

**Influence of nonlocal response from the structural periodicity on the Cherenkov threshold.**

We begin with the analysis of the influence of nonlocal response from the structural periodicity on the Cherenkov threshold in Fig. 1. That is, we consider the case that the nonlocal response from metals is negligible, by setting the thickness of metal $d_1$ much larger than the Tomas-Fermi screening length $\lambda_{\mathrm{TF}}$. The value of $\lambda_{\mathrm{TF}} \approx \frac{\beta}{\omega_p}$ is generally a few angstrom, which describes the spatial extent of surface charge densities for metals (*40*). Here we set $d_1 = 10$ nm $\gg \lambda_{\mathrm{TF}}$ and $P = 25$ nm in Fig. 1.

Figure 1 (B to E) shows the field distribution of free electron radiation in time domain. All field distributions here and below are calculated under the framework of classic electromagnetic wave theory; see methods in Supplementary Section 5. Under the local approximation, the structure in Fig. 1A has $\min(v_{\mathrm{th}}(\lambda))_{\mathrm{local}} = 0$ in the interested range of wavelength or the so-called threshold-free Cherenkov radiation in hyperbolic metamaterials; see radiation fields at two arbitrarily selected particle velocities in Fig. 1 (B and D).

In contrast, when under the nonlocal description, Cherenkov radiation inside the bulk layered structure appears if $v = 0.2c$ in Fig. 1C, but it disappears if $v = 0.05c$ in Fig. 1E. This provides a clear evidence that $\min(v_{\mathrm{th}}(\lambda))_{\mathrm{nonlocal}} > 0.05c$ in the interested range of wavelength [Fig. S4A]. Such a nonzero Cherenkov threshold originates from the fact that the nonlocal response from structural periodicity would prohibit the excitation of bulk Bloch modes with extremely large wavevectors. In addition, Cherenkov radiation of surface plasmons at the interface also shows up in Fig. 1 (C and E), when considering the realistic structure. From Fig. 1 (B to E), it is reasonable to argue that the local approximation fails to describe the Cherenkov threshold and the interaction between free electrons and metallic nanostructures, particularly when the free electron moves at a low speed.

**Influence of nonlocal response from the nonlocal electron screening in metals on the Cherenkov threshold.**

Next, we proceed to analyze in Fig. 2 the influence of nonlocal response from metals on the Cherenkov threshold. That is, we consider the case that the nonlocal response from metals is non-negligible, by setting $\beta \neq 0$ and $d_1$ comparable to $\lambda_{\text{TF}}$. Figure 2A shows the Cherenkov threshold $v_{\text{th}}(\lambda_0)$ as a function of the pitch $P$ of unit cell. Here we choose the wavelength $\lambda_0$, e.g., $\lambda_0 = 1$ µm, at which the metallodielectric layered structure in Fig. 1A under the local approximation is a hyperbolic metamaterial. If $P$ decreases, $v_{\text{th}}(\lambda_0)$ decreases in Fig. 2A. Moreover, $\lim_{P \to 0} v_{\text{th}}(\lambda_0) = \beta$. In other words, the phenomenological nonlocal parameter $\beta$ determines the ultimate low-bound for the Cherenkov threshold in the layered structure, or we have $v_{\text{th}}(\lambda_0) \geq \beta$ for the arbitrary value of $P$. As such, the Cherenkov threshold in hyperbolic metamaterials is always nonzero since $\beta \neq 0$ for realistic metals. The underlying mechanism is that the nonlocal response from metals regularizes the broadband singularity of photonic density of states in hyperbolic metamaterials, since the nonlocal metal is equivalent to a composite material, comprising a finite thin dielectric layer on top of a local metal (*41*). Note that if we let $\beta = 0$, the nonlocal response of metals would be artificially neglected, and the Cherenkov threshold reduces to zero if $P \to 0$ in Fig. 2A.

Figure 2A also indicates that the nonlocality in artificial nanostructures may be exploited to engineer Cherenkov thresholds in a flexible way. If $P$ is sufficiently small (e.g., $P < 1$ nm), the nonlocal response of hyperbolic metamaterials mainly arises from nonlocal metals, so that $v_{\text{th}}(\lambda_0) \sim \beta$. If $P$ is sufficiently large ($P > 10$ nm), the nonlocal response of hyperbolic metamaterials is dictated by the structural periodicity. As a result, a convergent tendency of $v_{\text{th}}(\lambda_0)$ emerges for different values of $\beta$, and it is possible to have $v_{\text{th}}(\lambda_0) \gg \beta$. If $1 \text{ nm} < P < 10 \text{ nm}$, $v_{\text{th}}(\lambda_0)$ is determined both by nonlocal metals and the structural periodicity.

To highlight the influence of nonlocal metals on Cherenkov radiation inside hyperbolic metamaterials, Fig. 2 (B to E) shows the free electron radiation in time domain, by setting $d_1 = 0.8$ nm and $P = 2$ nm. In Fig. 2 (B to E), $v = 0.015c$ is chosen; three different values of $\beta$ are used, which are 0, $c/300$ and $c/30$, respectively, in Fig. 2 (C to E); Fig. 2B is calculated under the local approximation. Cherenkov radiation inside hyperbolic metamaterials emerges in Fig. 2 (C and D) since $v > \beta$, but it disappears in Fig. 2E since $v < \beta$. These results numerically verify the nonzero Cherenkov threshold for realistic hyperbolic metamaterials. On the other hand, Fig. 2E shows the absence of Cherenkov radiation of surface plasmons at the interface if the nonlocal response of hyperbolic metamaterials is mainly induced by nonlocal metals. Such a result is distinct from Fig. 1E. In addition, the emitted fields of Cherenkov radiation in Fig. 2 (C and D) are similar with that in Fig. 2B, except for fields inside the triangular region highlighted in Fig. 2 (B to D). The difference comes from that the radiation spectrum of Cherenkov radiation is sensitive to $\beta$, since $v_{\text{th}}(\lambda)$ is dependent on both $\beta$ and $\lambda$.

**Dissipated power emitted by a swift electron in hyperbolic metamaterials.**

To facilitate the experimental observation of the revealed nonlocal Cherenkov threshold in hyperbolic metamaterials, Fig. 3 shows the comparison of dissipated powers $P_0$ for the moving electron in Fig. 1A by using the local and nonlocal descriptions. The total dissipated power is equal to the integration of energy loss spectrum $G$ of over the interested range of wavelength; see methods in Supplementary Section 7. From Fig. 3, the nonlocality induced from the structural periodicity has a strong influence on the dissipated power. In contrast, the nonlocal electron screening in metals would affect the dissipated power only when the pitch of unit cell is small (e.g., $P \leq 2$ nm in Fig. 3). As a clear signature for the nonlocal Cherenkov threshold, if $v$ decreases to a sufficiently small value (e.g., $v < 0.1c$ for $P = 10$ nm in Fig. 3), the dissipated power would dramatically drop, due to the disappearance of radiation channels in more frequencies. In addition, there is a large discrepancy between the dissipated powers from the local and nonlocal calculations in Fig. 3, indicating the local approximation is not accurate for the calculation of dissipated power.

**Influence of longitudinal modes on the Cherenkov threshold around epsilon-near-zero frequencies.**

Now recall the in-plane permittivity $\varepsilon_{\parallel}$ of the effective uniaxial material for the layered structure in Fig. 1A. Note that around the frequency of $\varepsilon_{\parallel} \to 0$, the longitudinal modes will show up with the consideration of nonlocality, in addition to the transverse modes (*35*). We then proceed to discuss the influence of longitudinal modes on the Cherenkov threshold in Fig. 4. In Fig. 4A, we let $\lambda_0 = 0.395$ μm, at which $\text{Re}(\varepsilon_{\parallel}) \to 0^+$ and $\text{Re}(\varepsilon_{\perp}) > 0$ [Fig. S1]. We reveal in Fig. 4A that the longitudinal mode has a much lower Cherenkov threshold than the transverse mode. Such an emerging phenomenon can be understood as follows.

On the one hand, for transverse modes, their isofrequency contour at $\lambda_0$ is elliptical [Fig. S1], since $\text{Re}(\varepsilon_{\perp}) \cdot \text{Re}(\varepsilon_{\parallel}) > 0$. The layered structure in Fig. 1A at $\lambda_0$ is then denoted as the elliptical metamaterial below. The elliptical isofrequency contour, whether calculated under the local or nonlocal descriptions [Fig. S3], would generally lead to a large Cherenkov threshold for transverse modes; for example, we always have $v_{\text{th}}^T(\lambda_0) > 0.2c$ in Fig. 4A. On the other hand, for longitudinal modes, the corresponding Cherenkov threshold $v_{\text{th}}^L(\lambda_0)$ is sensitive to the nonlocal response induced both by nonlocal metals and the structural periodicity in Fig. 4A. In short, the features of Cherenkov threshold for longitudinal modes inside elliptical metamaterials in Fig. 4A are similar with that for transverse modes inside hyperbolic metamaterials in Fig. 2A.

Since $v_{\text{th}}^L(\lambda_0) < v_{\text{th}}^T(\lambda_0)$ for the arbitrary value of *P* in Fig. 4A, it is then reasonable to argue that the consideration of nonlocality would largely *decrease* the Cherenkov threshold for elliptical metamaterials at epsilon-near-zero frequencies. Such a feature is distinct from hyperbolic metamaterials, in which the consideration of nonlocality would instead *increase* the Cherenkov threshold.

Figure 4 (B and C) shows the field distribution of free electron radiation at $\lambda_0$. The longitudinal mode does not show up in the local approximation [Fig. 4B], while it is excited by free electrons under the nonlocal description [Fig. 4C]. These results numerically verify the invalidity of local approximation in describing Cherenkov radiation inside elliptical

metamaterials at epsilon-near-zero frequencies. Figure 4 (D and E) shows the radiation spectrum of excited longitudinal modes around $\lambda_0$. Figure 4D shows that the excitation of longitudinal modes is sensitive to the particle velocity $v$ and the wavelength $\lambda$. To be specific, if $v$ is fixed, the longitudinal mode can be efficiently excited only within a narrow range of wavelength, although they exist in a relative wide range of wavelength [Fig. 4D]. Meanwhile, the longitudinal mode at a certain wavelength can be efficiently excited only for a certain range of $v$ [Fig. 4D]. Figure 4E indicates that the efficient excitation of longitudinal and transverse modes at the same wavelength requires different values of $v$. For example, at $\lambda_0$, the peak for the efficient excitation of longitudinal modes appears at $v = 0.2c$, while the peak for transverse modes is at $v = 0.6c$. With a proper design, the values of these two peaks are in the same order of magnitude. These features might facilitate the potential observations of Cherenkov radiation of longitudinal modes at epsilon-near-zero frequencies.

**DISCUSSION**

In summary, we for the first time theoretically prove the *non-existence* of threshold-free Cherenkov radiation in hyperbolic metamaterials. Instead, the realistic nonlocality would always enable an ultimate lower-bound velocity cutoff (whose value is comparable to the Fermi velocity of electrons in metals) inside metallodielectric layered structures, including hyperbolic and elliptical metamaterials. Moreover, we reveal that while the nonlocality would increase the value of Cherenkov threshold in hyperbolic metamaterials, it would, in turn, decrease the value of Cherenkov threshold in elliptical metamaterials. In addition, although the hydrodynamic model is a very approximate approach in treating the nonlocal and quantum effects in metallic nanostructures, we emphasize that our theoretical predictions of nonlocal Cherenkov threshold is universal regardless of which quantum models used. This is because the observed threshold behaviors of Cherenkov radiation in our configuration only rely on the reduced photonic density of states and the appearance of longitudinal modes. Previous experimental results have verified that the presented approaches in this work (i.e., the hydrodynamic model and the realistic periodic structure) can precisely anticipate these two phenomena induced by the nonlocality (*40, 42*). In short, the nonlocality could significantly re-shape Cherenkov radiation in complex media, including the Cherenkov threshold, and it provides a valuable way to flexibly engineer the Cherenkov threshold. Our work thus provides an important theoretical guidance for many practical applications of Cherenkov radiation, such as the design of novel integrated free-electron radiation sources on chip.

**MATERIALS AND METHODS**

To reveal the impacts of nonlocality, we compare Cherenkov radiation in three configurations, i.e., homogeneous hyperbolic metamaterial, layered structure with the consideration of nonlocality only from the structural periodicity, and layered structure with the consideration of nonlocality from both the structural periodicity and the electron screening in metals. In the first configuration, we treat the hyperbolic material as a homogeneous medium, with the effective permittivity given by Maxwell Garnet approximation. In the last two configurations, the permittivities are explicitly defined for each region of the layered structures. Without considering the nonlocal electron screening in metals, we calculate the fields in all region using potentials with decomposition of the magnetic field and the transverse electric field. When the nonlocal electron screening in metals is considered, we include an additional longitudinal electric field with hydrodynamic model. The transmission and reflection coefficients of the transverse field or the longitudinal field are analytically solved by employing the method of transfer matrix. In addition, Cherenkov thresholds in last two configurations are analytically determined from the corresponding isofrequency contours when the material loss is artificially neglected. Finally, we evaluate the dissipated power emitted by the swift electron using the power dissipation formula.

**Acknowledgements**

**Funding:** Y. Luo was sponsored in part by Singapore Ministry of Education under Grant No. MOE2018-T2-2-189 (S), 2017-T1-001-239 (RG91/17 (S)); A*Star AME programmatic Grant No. A18A7b0058. B. Zhang was partially supported by Nanyang Technological University for NAP Start-Up Grant and the Singapore Ministry of Education (Grant No. MOE2018-T2-1-022 (S), MOE2016-T3-1-006 and Tier 1 RG174/16 (S)). **Author contributions:** All authors contributed extensively to the work presented in this paper. Y.L. and H.H. conceived the research. H.H. performed the calculation. X.L., J.Z., D.L, P.G., B.Z. and Y.L. contributed insight and discussion on the results. H.H., X.L., Y.L., and B.Z. wrote the paper. Y.L., X.L. and B.Z. supervised the project. **Competing interests:** The authors declare that they have no competing financial interests. **Data and materials availability.** All data needed to evaluate the conclusions are present in the paper and/or the Supplementary Materials. Additional data related to this paper may be requested from the authors.

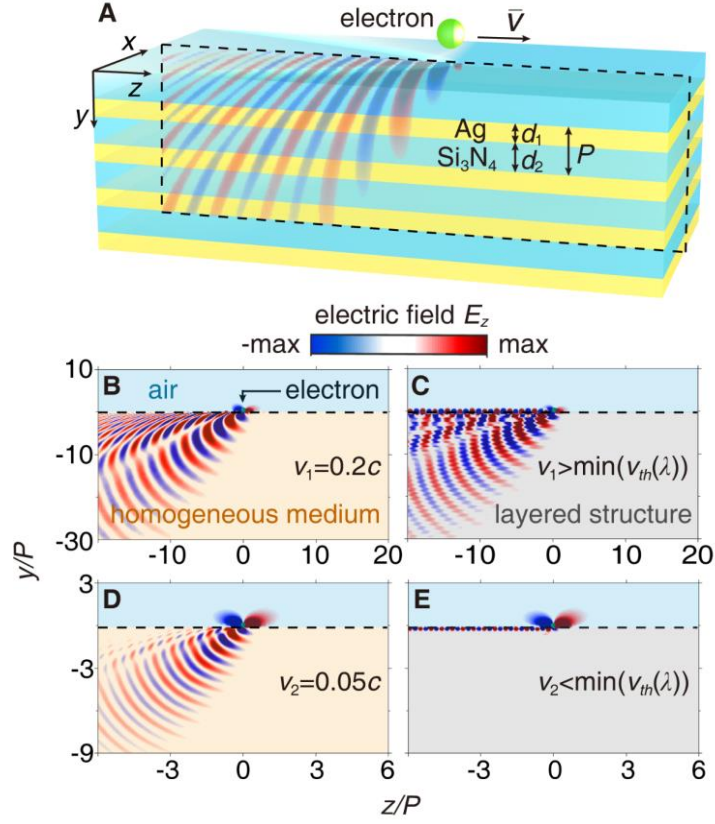

**Fig. 1. Influence of nonlocal response from the structural periodicity on the Cherenkov threshold**. (**A**) Structural schematic. The swift free electron has a velocity of $\bar{v} = \hat{z}v$. (**B-E**) Field distribution of free electron radiation in time domain. The realistic layered structure is considered in (C and E), while it is treated as a homogeneous uniaxial material in (B and D) according to the classical local description. The realistic nanostructure in A has a wavelength-dependent Cherenkov threshold $v_{\text{th}}(\lambda)$, namely the particle velocity threshold required for the emergence of Cherenkov radiation. Cherenkov radiation appears if $v = v_1 > \min(v_{\text{th}}(\lambda))$ in C but disappears if $v = v_2 < \min(v_{\text{th}}(\lambda))$ in E. Other parameters are: the pitch of unit cell $P = d_1 + d_2 = 25$ nm; for all studied cases here and below, $d_1/P = 0.4$, and the interested range of wavelength in free space is $\lambda \in [0.1\ 10]$ μm.

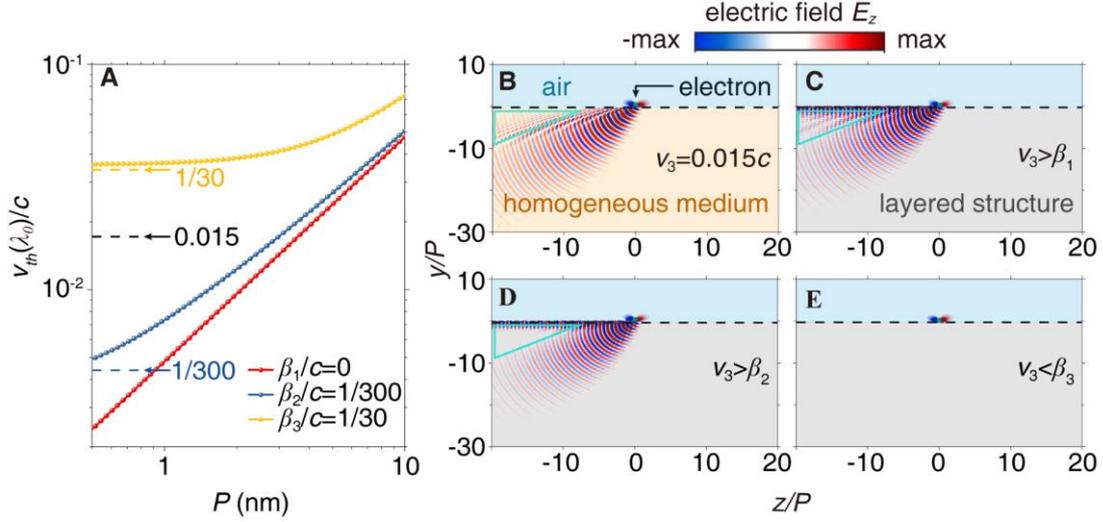

**Fig. 2. Influence of nonlocal response from the nonlocal electron screening in metals on the Cherenkov threshold.** (**A**) Cherenkov threshold $v_{\text{th}}(\lambda_0)$ as a function of the pitch $P$ of unit cell, where $\lambda_0 = 1$ μm. $\beta$ is the phenomenological nonlocal parameter for metals. For realistic metals, $\beta \neq 0$. The metallodielectric layered structure in Fig. 1A under the local approximation is the hyperbolic metamaterial at $\lambda_0$. Due to the nonlocal metals, we always have $v_{\text{th}}(\lambda_0) \geq \beta$. (**B-E**) Field distributions of free electron radiation in time domain. The realistic layered structure in Fig. 1A is adopted for (C to E), while it is replaced by an effective homogeneous uniaxial material for B according to the local approximation. Cherenkov radiation appears if $v_3 > \beta$ in (C and D) but disappears if $v_3 < \beta$ in E. Other parameters are $P = 2$ nm, and $v = v_3 = 0.015c$ in (B to E).

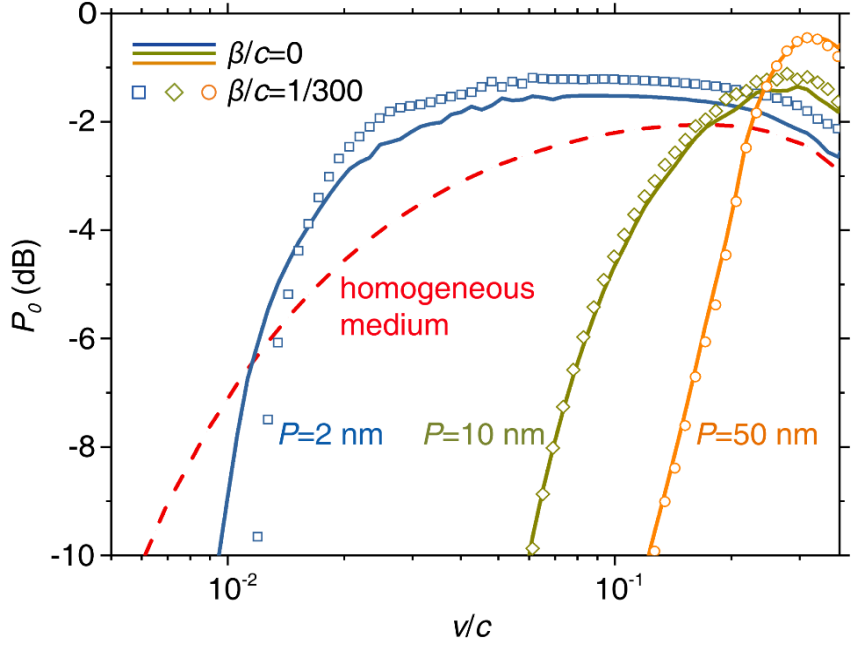

**Fig. 3. Influence of the nonlocality on the dissipated power of a swift electron moving above metallodielectric layered structures.** For comparison, metallodielectric layered structure in Fig. 1A is treated by three different ways. For the red dashed line, the layered structure is treated as an effective homogeneous uniaxial medium, according to the classical local description. For the solid lines, the nonlocal response from the structural periodicity is considered by setting the nonlocal parameter $\beta = 0$. For the data indicated by hollow symbols, the nonlocal responses both from the structural periodicity and from the nonlocal electron screening in metals (e.g., $\beta/c = 1/300$) are taken into consideration. When $d_1/P$ is fixed, the dissipated power calculated from the classical local description is independent of $P$.

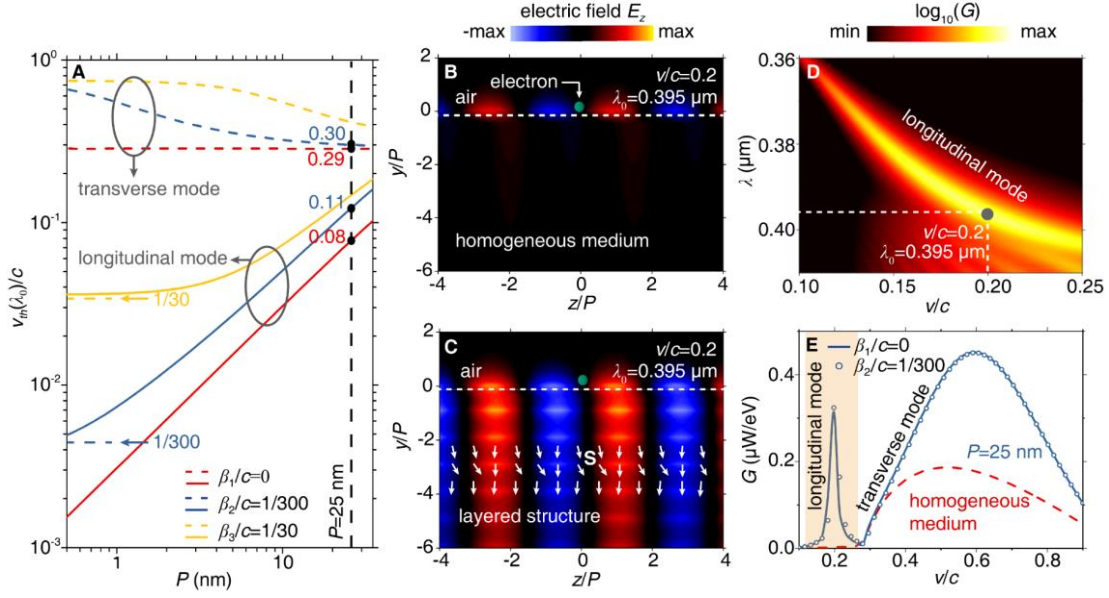

**Fig. 4. Influence of longitudinal modes on the Cherenkov threshold around the epsilon-near-zero frequency.** At the wavelength around $\lambda_0 = 0.395$ μm, the metallodielectric layered structure in Fig. 1A under the local approximation has an elliptical isofrequency contour and one component of its effective permittivity close to zero. Then the realistic layered structure in Fig. 1A supports both transverse and longitudinal modes around $\lambda_0$. (**A**) Cherenkov thresholds of longitudinal and transverse modes vs the pitch $P$ of unit cell at $\lambda_0$, with the consideration of nonlocality. The results calculated from the local approximation is the same as the red dashed line. (**B, C**) Field distributions of free electron radiation at $\lambda_0$. The longitudinal mode appears in C if the realistic layered structure in Fig. 1A is considered, but it vanishes in B if the local approximation is used. (**D**) Electron energy loss spectrum $G$ as a function of wavelength $\lambda$ and particle velocity $v$. (**E**) Electron energy loss spectrum $G$ vs $v$ at $\lambda_0$. For (B to E), $P = 25$ nm.